\documentclass[amsmath,amssymb,aps,prfluids]{revtex4-2}
\usepackage{graphicx}% Include figure files
\usepackage{bm} % bold math
\usepackage[]{tikz}

\begin{document}

\title{On the instability of the magnetohydrodynamic pipe flow subject to a transverse magnetic field}

\author{Yelyzaveta Velizhanina}
 \email[]{yelyzaveta.velizhanina@ulb.be}
\author{Bernard Knaepen}
 \email[]{bernard.knaepen@ulb.be}
\affiliation{ Physique des Systèmes Dynamiques, Faculté des Sciences, Université Libre de Bruxelles, Boulevard du Triomphe CP231, 1050 Ixelles, Belgium}

\date{\today}

\begin{abstract}
The linear stability of a fully-developed liquid-metal MHD pipe flow subject to a transverse
magnetic field is studied numerically. Because of the lack of axial symmetry in the mean velocity profile, we need to perform a BiGlobal stability analysis. For that purpose, we develop a two-dimensional complex eigenvalue solver relying on a Chebyshev-Fourier collocation method in physical space.
By performing an extensive parametric study, we show that in contrast to Hagen-Poiseuille
flow known to be linearly stable for all Reynolds numbers, the MHD pipe flow with transverse
magnetic field is unstable to
three-dimensional disturbances at sufficiently high values of the Hartmann
number and wall conductance ratio. The instability observed in this regime is attributed to the presence of velocity overspeeds in the so-called Roberts layers and the corresponding inflection points in the mean velocity profile. The nature and characteristics of the most unstable modes are investigated, and we show that they vary significantly depending on the wall conductance ratio. A major result of this paper is that the global critical Reynolds number for the MHD pipe flow with transverse magnetic field is $Re=45230$, and it occurs
for a perfectly conducting pipe wall and the Hartmann number $Ha=19.7$.
\end{abstract}

\maketitle

\section{Introduction}\label{section:introduction}

Linear stability theory suggests that all three-dimensional perturbations
of small amplitude superposed on Hagen-Poiseuille flow -- the flow of
viscous incompressible fluid in a circular pipe -- decay after sufficiently long times \cite{Lessen1968,Meseguer2003}. However, the experiments show that the flow can undergo a laminar-turbulent transition at a finite Reynolds number \cite{Reynolds1883,Wygnanski1973}.
In other words, the observed transition is subcritical, \textit{i.e.}
occurs for a Reynolds number below that predicted by the modal linear stability analysis. Similar to other shear flows, a likely scenario of transition to turbulence in the pipe consists in destabilization of the flow by finite amplitude perturbations that arise due to the large transient growth of initially small three-dimensional disturbances \cite{Schmid2001,Pringle2010}. The most amplified initial disturbance of certain critical amplitude may then lead to what is known as an edge state -- a solution of the equations of motion that will neither relaminarize nor trigger turbulence \cite{Pringle2015,Avila2023}. This provides an important insight into the matter and allows finding effective ways to control the occurrence of instabilities.

Despite the significant advancement in the understanding of the instability mechanism in
Hagen-Poiseuille flow, little is known about the onset of turbulence in a liquid
metal flow in a circular pipe subject to an applied magnetic field. Since the
classical experiments of Hartmann \& Lazarus \cite{Hartmann1937}, it is well known
that an applied magnetic field can stabilize an otherwise unstable flow. This phenomenon
is exploited in numerous applications, such as semiconductor crystal growth
\cite{Chen2008}, steel casting \cite{Chaudhary2012} and flow control \cite{Tsinober1990}.
The mechanism behind this stabilization is Joule damping which occurs as a result of the creation of electric currents induced by the motion of the fluid in the presence of an applied magnetic field.

However, in certain configurations, an applied magnetic field can also have a destabilizing effect. In the case of the duct flow with a uniform transverse magnetic field, the linear stability analysis was performed by Priede {\em et al.} \cite{Priede2010,Priede2012,Priede2016} who showed that the flow is unstable in the
presence of electrically conducting walls at sufficiently high Hartmann numbers. The source of instability originates from the modification of the mean velocity profile and the presence of sidewall velocity jets away from the centerline of the duct. This effect is particularly pronounced in Hunt's flow, which is characterized by a pair of perfectly conducting walls and a pair of insulating walls in the direction parallel to the magnetic field. In the case of insulating walls, the duct flow remains linearly stable and its subcritical transition through transient growth was studied by Krasnov {\em et al.} \cite{Krasnov2010}. The MHD duct flow in the non-linear regime has also received significant attention (see \cite{Zikanov2014} for a review) and recently Blishchick {\em et al.} have highlighted the influence of the wall conductivity on the flow regime \cite{Blishchik2021}. Their observations are consistent with the linear computations of Priede {\em at al.} who highlighted the important role played by sidewall velocity jets for certain combinations of parameters.

To the best of our knowledge, no detailed linear stability analysis of the MHD pipe flow with a transverse magnetic field has been performed, although several authors have considered the stability of pipe flows subject to natural convection and an applied magnetic field \cite{Zikanov2013,Hu2022}. More relevant to the present study is the work by {\AA}kerstedt \cite{Akerstedt1995} who analyzed the stabilizing effect of a longitudinal magnetic field on the MHD pipe
flow with a perfectly conducting wall and also performed a transient growth analysis in that case. Some results concerning the relaminarization of the MHD pipe flow and its dynamics close to the transitional regime are also available, for example, in \cite{Gardner1971,Zikanov2014,Moriconi2020}.

As in a duct, a transverse magnetic field has potentially two effects on the stability of the MHD pipe flow: firstly, it can stabilize the flow by introducing extra Joule dissipation; secondly, it can destabilize it by modifying the mean velocity profile. In a pipe, a transverse magnetic field systematically elongates the flow along its direction and induces thin Hartmann layers in regions where the wall is normal to the magnetic field.
In the regions where the wall is parallel to the magnetic field, another type of boundary layer develops, known as a Roberts layer \cite{Ihara1967, Gold1962,Shercliff1956}. If the pipe's wall is electrically insulating, the velocity profile in these layers decay monotonically with the distance from the wall. On the contrary, at sufficiently high wall conductivity and Hartmann numbers, regions of velocity overspeed with inflection points
may appear \cite{Samad1981,Vantieghem2009}. The situation is thus very similar to the one observed in the duct, and it is therefore interesting to assess whether modal instability is possible in the MHD pipe.

In this paper, we address the above question using BiGlobal stability analysis -- a notion
established in \cite{Theofilis2011} to describe a linear
stability analysis in a three-dimensional domain with two inhomogeneous and one
homogeneous directions. This allows us to formulate the stability of the flow
as a two-dimensional eigenvalue
problem which depends on the Reynolds number, the intensity of the applied magnetic field and the ratio of electric
conductivity of the wall to that of the fluid.
The Reynolds and Hartmann numbers considered in the parametric study range
from $10^3$ to $10^6$ and $0$ to $60$, respectively, while the conductance ratio is varied from $0$ to $\infty$.

The paper is organized as follows. In Section \ref{section:problem-formulation}, we
present the physical model and the formulation of the problem. Section \ref{section:numerical-method}
is dedicated to the description of the numerical method and to its validation. Our results on the
global stability of the MHD pipe flow in a transverse magnetic field are
then presented in Section \ref{section:results} while Section \ref{section:conclusions} contains the conclusions of the present study.

\section{Problem formulation}\label{section:problem-formulation}
\begin{figure}[ht!]
  \includegraphics[width=0.6\textwidth]{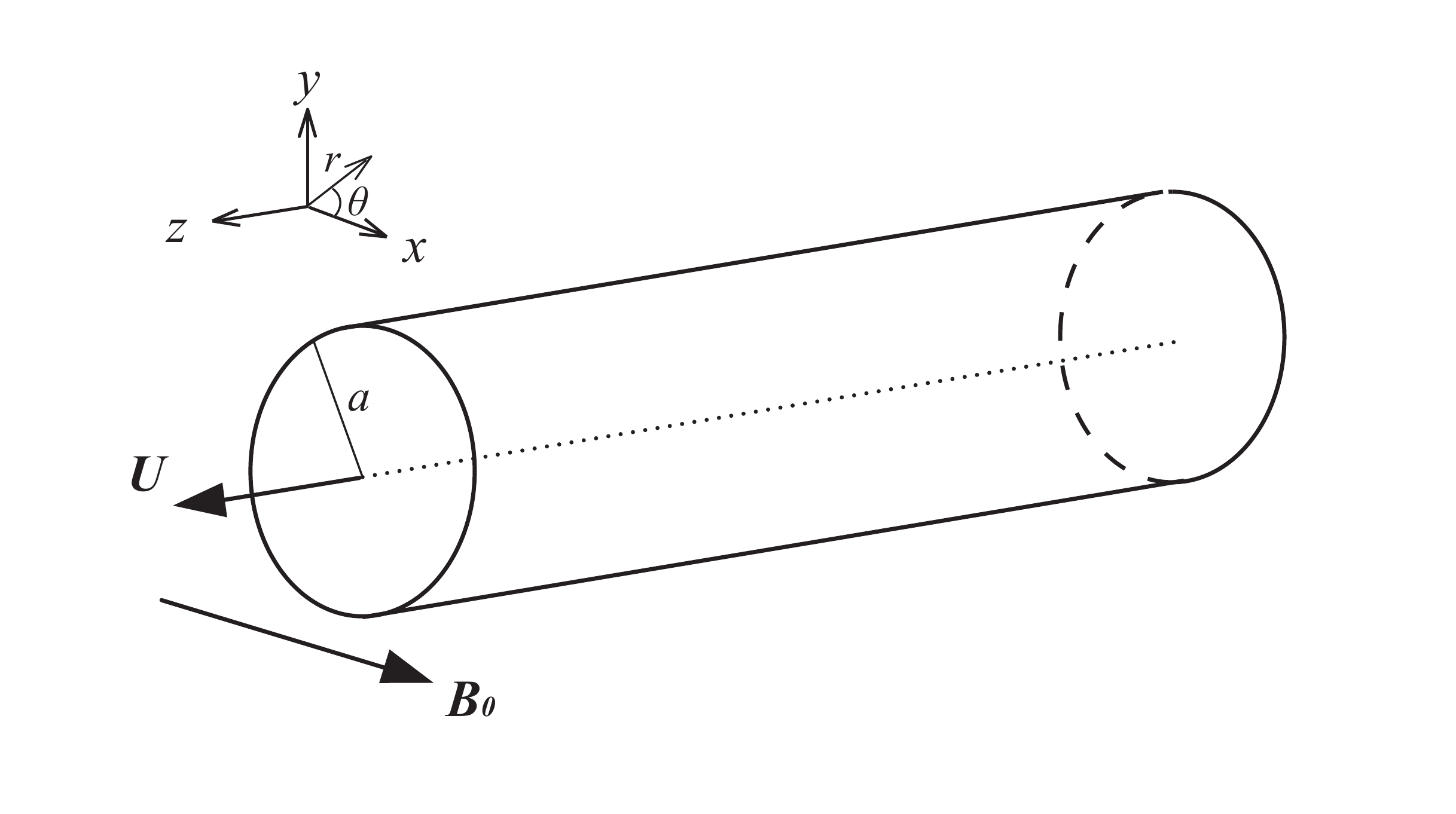}
  \caption{Schematic representation of the MHD pipe flow subject to a uniform
  transverse magnetic field.}
  \label{fig:geom}
\end{figure}

%\subsection{Governing equations}
We consider the flow of an incompressible viscous electrically
conducting fluid in a circular pipe of radius $a$ subject to a transverse uniform
magnetic field $\bm{B}_0$ (see Fig.~\ref{fig:geom}).

We focus on liquid-metal flows and assume that the quasi-static approximation is applicable. In this framework the induced magnetic field is negligible compared to
the applied magnetic field and the characteristic timescale of magnetic
diffusion is small compared to the other relevant timescales. The velocity $\bm{u}$
and the pressure $p$ are then modeled by the quasi-static MHD equations \cite{Muller2001},
\begin{subequations}\label{eq:dim-ns}
  \begin{eqnarray}
    \nabla \cdot \bm{u} = 0,\\
    \frac{\partial \bm{u}}{\partial t} + \left( \bm{u} \cdot \nabla \right) \bm{u} = -\frac{1}{\rho}\nabla p + \nu\nabla^2 \bm{u} + \frac{1}{\rho} \left( \bm{j} \times \bm{B}_0 \right),
  \end{eqnarray}
\end{subequations}
where $\rho$, $\nu$ and $\sigma$ are, respectively, the density, the kinematic viscosity and the electrical conductivity of the fluid. The induced electric current $\bm{j}$ obeys Ohm's law
for a moving conductor,
\begin{equation}
  \bm{j} = \sigma\left( -\nabla \phi + \bm{u}\times\bm{B}_0 \right),\label{eq:ohm}
\end{equation}
and satisfies the charge conservation equation,
\begin{equation}
  \nabla\cdot\bm{j} = 0\label{eq:dim-charge-cons}.
\end{equation}
The electric potential $\phi$ appearing in Eq.~(\ref{eq:ohm}) may be eliminated by applying the curl operator on $\bm{j}$.

At the pipe's wall, $\bm{u}$ satisfies the no-slip and impermeability
conditions,
\begin{equation}
  \bm{u} = 0 \quad \textrm{at}\,\,\, r = a.\label{eq:dim-bc-u}
\end{equation}
The boundary condition for $\bm{j}$ is obtained under the `thin-wall'
assumption, valid when the thickness of the wall $t_w$ is much smaller
than the radius of the pipe $a$ \cite{Muller2001},
\begin{equation}
  \bm{j}\cdot\bm{n} - \nabla_\tau \cdot \left( \frac{t_w \sigma_w}{\sigma} \bm{j}_{\tau} \right) = 0 \quad \textrm{at}\,\,\, r = a,\label{eq:dim-bc-current}
\end{equation}
where $\sigma_w$ is the conductivity of the wall, $\bm{n}$ is the outward wall normal, $\bm{j}_\tau = \bm{j} - \left( \bm{j}\cdot\bm{n} \right)\bm{n}$
and $\nabla_\tau = \nabla - (\partial/\partial n)\bm{n}$.
Note that this boundary condition reduces to $\bm{j}\cdot\bm{n} = 0$ for a perfectly insulating wall and $\nabla_\tau \cdot \bm{j}_\tau = 0$ for a perfectly conducting wall.

The above equations are made non-dimensional using the following transformations:
\begin{equation}
  \begin{matrix}\label{eq:nondim-vars}
    & \bm{x} \to a\bm{x}, & t \to a U_0^{-1}t, & p \to \rho U_0^2 p,\\
    & \bm{u} \to U_0 \bm{u}, & \bm{B}_0 \to B_0 \bm{1}_{B_0}, & \bm{j}\to \sigma U_0 B_0 \bm{j},
  \end{matrix}
\end{equation}
where the characteristic velocity scale $U_0$ is the maximum base flow velocity (see below),
$B_0=\vert \bm{B}_0 \vert$ is the magnitude of the applied magnetic field and
$\bm{1}_{B_0}$ is a unit vector, aligned with $\bm{B}_0$. Note that in
the cylindrical coordinates $\left( r, \theta, z \right)$ depicted in Fig.~\ref{fig:geom},
$\bm{1}_{B_0} = \cos{\theta}\bm{1}_r - \sin{\theta}\bm{1}_\theta$.

In terms of the non-dimensional variables (\ref{eq:nondim-vars}), the
system given by Eqs.~(\ref{eq:dim-ns})-(\ref{eq:ohm})-(\ref{eq:dim-charge-cons}) and the boundary conditions (\ref{eq:dim-bc-u})-(\ref{eq:dim-bc-current}) read
\begin{subequations}\label{eq:nondim-mhd-eqs}
  \begin{eqnarray}
    \nabla \cdot \bm{u} = 0,\\
    \frac{\partial \bm{u}}{\partial t} + \left( \bm{u} \cdot \nabla \right) \bm{u} = -\nabla p + \frac{1}{Re}\nabla^2 \bm{u} + \frac{Ha^2}{Re} \left( \bm{j} \times \bm{1}_{B_0} \right),\\
    \nabla\times\bm{j} = \left( \bm{1}_{B_0}\cdot\nabla \right)\bm{u},\\
    \nabla\cdot\bm{j} = 0,\label{eq:divfree}
  \end{eqnarray}
\end{subequations}
with boundary conditions,
\begin{subequations}
  \begin{eqnarray}
    \bm{u} = 0 \quad & \textrm{at}\,\,\, r = 1,\label{eq:nondim-no-slip}\\
    \bm{j}\cdot\bm{n} - \nabla_\tau \cdot \left( \chi \bm{j}_{\tau} \right) = 0 \quad & \textrm{at}\,\,\, r = 1.\label{eq:nondim-thin-wall}
  \end{eqnarray}
\end{subequations}
The non-dimensional parameters $Re$, $Ha$ and $\chi$ appearing in this set of equations are, respectively,
the Reynolds number, the Hartmann number and the wall conductance ratio:
\begin{equation}
  Re = \frac{U_0 a}{\nu}, \quad Ha = B_0 a\sqrt{\frac{\sigma}{\rho\nu}}, \quad \chi = \frac{t_w \sigma_w}{a\sigma}.
\end{equation}

To address the linear stability of the fully developed MHD pipe
flow in the presence of a transverse magnetic field we decompose the instantaneous variables $\bm{u}$, $p$ and $\bm{j}$ into a base state $(\bm{U}, P, \bm{J})$ and infinitesimal perturbations $(\bm{u}', p', \bm{j}')$ according to
\begin{equation}
  \bm{u} = \bm{U} + \bm{u}', \quad p = P +  p', \quad \bm{j} = \bm{J} + \bm{j}'.\label{eq:full-mhd-vars}
\end{equation}
In cylindrical coordinates, the base state is given by
\begin{equation}
  \bm{U} = U\left( r, \theta \right)\bm{1}_z, \quad P = P\left( r, \theta, z \right), \quad \bm{J} = J_r\left( r, \theta \right)\bm{1}_r + J_\theta\left( r, \theta \right)\bm{1}_\theta.\label{eq:base-flow-vars}
\end{equation}
To compute it, we solve the following system of scalar equations,
\begin{subequations}\label{eq:base-flow}
  \begin{eqnarray}
    \nabla^2 U + Ha^2 \left( \cos{\theta}\frac{\partial}{\partial r} - \frac{\sin{\theta}}{r}\frac{\partial}{\partial \theta} \right) M_z = Re K,\\
    \nabla^2 M_z + \left( \cos{\theta}\frac{\partial}{\partial r} - \frac{\sin{\theta}}{r}\frac{\partial}{\partial \theta} \right) U = 0,
  \end{eqnarray}
\end{subequations}
obtained after substituting $(\bm{U}, P, \bm{J})$ into Eqs.~(\ref{eq:nondim-mhd-eqs}) and using Eq.~(\ref{eq:divfree}) to express the mean electric current as $\bm{J}=\nabla \times \bm{M}$. Here $K = \partial P / \partial z$ is the streamwise pressure drop.

The flow disturbances $(\bm{u}', p', \bm{j}')$ are then decomposed into the following normal modes:
\begin{equation}\label{eq:waves}
  \left\{ \bm{u}', p', \bm{j}' \right\}
  \left( r, \theta, z, t \right) = \left\{ \hat{\bm{u}}, \hat{p}, \hat{\bm{\jmath}} \right\}
  \left( r, \theta \right) e^{i\alpha \left( z - ct \right)},
\end{equation}
where $\alpha c = \alpha c_r + i\alpha c_i$.
If $c_i$ is negative, the temporal growth rate of the perturbation is negative, and we say that the flow is linearly stable, whereas if $c_i$
is positive, the flow is linearly unstable.
Finally, by substituting this normal mode decomposition into Eqs.~(\ref{eq:nondim-mhd-eqs}) and linearizing them around the steady laminar flow (\ref{eq:base-flow-vars}), we obtain the following two-dimensional perturbation equations:
\begin{subequations}\label{eq:pert}
  \begin{eqnarray}
    \nabla \cdot \hat{\bm{u}} = 0,\label{eq:pert-1}\\
    i \alpha c \hat{\bm{u}} = i\alpha U\hat{\bm{u}} + \left( \hat{u}_r\frac{\partial}{\partial r} + \frac{\hat{u}_\theta}{r}\frac{\partial}{\partial\theta} \right) \bm{U} + \nabla \hat{p} - \frac{1}{Re}\nabla^2 \hat{\bm{u}} - \frac{Ha^2}{Re}\left( \hat{\bm{\jmath}} \times \bm{1}_{B_0} \right),\label{eq:pert-2}\\
    \nabla \times \hat{\bm{\jmath}} = \left( \bm{1}_{B_0} \cdot \nabla \right) \hat{\bm{u}}.\label{eq:pert-3}\\
    \nabla \cdot \hat{\bm{\jmath}} = 0,\label{eq:pert-4}
  \end{eqnarray}
\end{subequations}
where $\displaystyle\nabla = \frac{\partial}{\partial r}\bm{1}_r + \frac{1}{r}\frac{\partial}{\partial\theta}\bm{1}_\theta + i\alpha \bm{1}_z$. According to Theofilis \cite{Theofilis2011}, the above system constitutes a BiGlobal stability problem as the unknowns depend on the two coordinates $r$ and $\theta$.

\section{Numerical method}\label{section:numerical-method}
To discretize the base flow equations (\ref{eq:base-flow}) and the perturbation equations
(\ref{eq:pert}) in $r$ and $\theta$ we use a combination of
Chebyshev and Fourier spectral collocation methods. The
unknowns are therefore expanded over suitable cardinal functions, and we look for
solutions in physical space $\bm{x} = \left( r, \theta \right)$ \cite{Boyd2000}.
In order to reduce the clustering of the grid points near the origin of the pipe
and to eliminate the coordinate singularity at $r=0$, we exploit the so-called
`rotate-and-reflect' symmetry \cite{Trefethen2000}. Any scalar field
$f\left( r, \theta, z, t \right)$ or vector field $\bm{g}\left( r, \theta, z, t \right)$ is first discretized in the redundant virtual domain $\left[-1, 1 \right] \times \left[ -\pi / 2, \pi / 2 \right]$ at $\tilde N = \tilde N_r N_\theta$ grid points, where $\tilde N_r$
and $N_\theta$ are, respectively, the number of Chebyshev and Fourier collocation
points. The following symmetries are then imposed:
\begin{subequations}
  \begin{eqnarray}
    f\left(r, \theta, z, t \right) = f\left(-r, \theta\pm\pi, z, t \right),\\
    g_r\left(r, \theta, z, t \right) = -g_r\left(-r, \theta\pm\pi, z, t \right),\\
    g_\theta\left(r, \theta, z, t \right) = -g_\theta\left(-r, \theta\pm\pi, z, t \right),\\
    g_z\left(r, \theta, z, t \right) = g_z\left(-r, \theta\pm\pi, z, t \right),
  \end{eqnarray}
\end{subequations}
and in such a way, the original polar domain $\left[ 0, 1 \right] \times \left[ -\pi / 2, \pi / 2 \right]$
is discretized at $N = N_r N_\theta$ grid points, where $N_r = \tilde N_r /2$.

Evaluating the continuous flow quantities
$\hat{\bm{u}}\left( r, \theta \right)$, $\hat{p}\left( r, \theta \right)$
and $\hat{\bm{\jmath}}\left( r, \theta \right)$ at the grid points yields the vectors
\begin{align}
\hat{\bm{U}} &= \left( \hat{u}_{r,0}, \dots ,\hat{u}_{r,N-1}, \hat{u}_{\theta,0}, \dots ,\hat{u}_{\theta,N-1}, \hat{u}_{z,0}, \dots , \hat{u}_{z,N-1} \right)^T,\\
\hat{P} &= \left( \hat{p}_0, \dots ,\hat{p}_{N-1} \right)^T, \\
\hat{\bm{J}} &= \left( \hat{\jmath}_{r,0}, \dots ,\hat{\jmath}_{r,N-1}, \hat{\jmath}_{\theta,0}, \dots ,\hat{\jmath}_{\theta,N-1}, \hat{\jmath}_{z,0}, \dots , \hat{\jmath}_{z,N-1} \right)^T,
\end{align}
with respective sizes $3N$, $N$ and $3N$. The discretized counterpart of the
system of equations (\ref{eq:pert}) then reads
\begin{subequations}\label{eq:disc-pert}
  \begin{eqnarray}
    F \hat{\bm{U}} = 0,\\
    i\alpha c \hat{\bm{U}} = S \hat{\bm{U}} + T \hat{P} + Q \hat{\bm{J}},\label{eq:disc-U-pert}\\
    G \hat{\bm{J}} = E\hat{\bm{U}}.\label{eq:disc-j-eq}
  \end{eqnarray}
\end{subequations}
Here $S$, $Q$, $G$ and $E$ are square matrices of size $3N$, whereas $F$ and
$T$ are rectangular matrices of sizes $N\times 3N$ and $3N\times N$, respectively.

Eq.~(\ref{eq:disc-j-eq}) is obtained by combining two linearly
independent components of Eq.~(\ref{eq:pert-3}) and Eq.~(\ref{eq:pert-4}).
The matrices $G$ and $E$ are thus defined as
\[
G=\left[\phantom{\begin{matrix}R D^{(1)}_\theta \\-D^{(1)}_r \\D^{(1)}_r + R\\R D^{(1)}_\theta \end{matrix}}
\right.\hspace{-4em}
\begin{matrix}
0 & -i\alpha I & R D^{(1)}_\theta  \\
i\alpha I & 0 & -D^{(1)}_r  \\
D^{(1)}_r + R & R D^{(1)}_\theta & i\alpha I \\
I & -\chi R D^{(1)}_\theta & -\chi i\alpha I
\end{matrix}
\hspace{-4em}
\left.\phantom{\begin{matrix}R D^{(1)}_\theta \\-D^{(1)}_r \\D^{(1)}_r + R\\R D^{(1)}_\theta \end{matrix}}\right]\hspace{-1em}
\begin{tabular}{l}
  $\scriptstyle \phantom{\begin{matrix} R D^{(1)}_\theta \ \end{matrix}} \hspace{-2.65em}\big\rbrace N$\\
  $\scriptstyle \phantom{\begin{matrix} R D^{(1)}_\theta\ \end{matrix}} \hspace{-2.65em}\big\rbrace N$\\
  $\scriptstyle \phantom{\begin{matrix} -D_r \ \end{matrix}} \hspace{-2em}\big\rbrace \left(N_r - 1\right)N_\theta$\\
  $\scriptstyle \phantom{\begin{matrix} R D^{(1)}_\theta\ \end{matrix}} \hspace{-2.65em}\big\rbrace N_\theta$\\
\end{tabular}\\
\begin{tabular}{lll}
  \hspace{-17.5em}\vspace{-6em}
  $\underbrace{\phantom{-R D^{(1)}_\theta}}_{N}\,\,$ & $\underbrace{\phantom{D^{(1)}_r + R}}_{N}\,\,$ & $\underbrace{\phantom{R D^{(1)}_\theta }}_{N}$
\end{tabular}
\]
and
\begin{equation}
  E = \begin{bmatrix}
    \cos{\Theta}D^{(1)}_r - \sin{\Theta}R D^{(1)}_\theta & \sin{\Theta}R & 0 \\
    -\sin{\Theta}R & \cos{\Theta}D^{(1)}_r - \sin{\Theta}R D^{(1)}_\theta & 0 \\
    0 & 0 & 0 \\
    0 & 0 & 0
  \end{bmatrix},
\end{equation}
where $I$ is
an identity matrix, $R=\text{diag}\left( r_j^{-1} \right)$,
$\Theta=\text{diag}\left( \theta_j \right)$ for $j=0,1,\dots N-1$,  and $D^{(n)}_r$ and $D^{(n)}_\theta$ are, respectively,
the $n$-th order differential matrices with respect to $r$ and $\theta$.
The electromagnetic boundary condition for $\hat{\bm{\jmath}}$ is taken
into account by replacing the lines of $G$ and $E$ corresponding
to Eq.~(\ref{eq:pert-4}) evaluated at $r=1$ with the discretized counterpart
of Eq.~(\ref{eq:nondim-thin-wall}). The latter is represented by a
matrix of size $N_\theta \times 3 N$.

If we eliminate $\hat{\bm{J}}$ from Eq.~(\ref{eq:disc-U-pert}) using Eq.~(\ref{eq:disc-j-eq}), we can rewrite our system of equations in terms
of the hydrodynamic variables $\hat{\bm{U}}$ and $\hat{P}$ only,
\begin{subequations}\label{eq:disc-pert-u-p}
  \begin{eqnarray}
    F \hat{\bm{U}} = 0,\\
    i\alpha c \hat{\bm{U}} = S' \hat{\bm{U}} + T \hat{P},\label{eq:disc-ns}
  \end{eqnarray}
\end{subequations}
with $S' = S + Q \left( G^{-1} E \right)$. Here $S$ represents the differential matrix
\[
S=\left[\phantom{\begin{matrix}i\alpha \mathcal{U} - Re^{-1}\left( D^2 - R^2 \right) \\i\alpha \mathcal{U} - Re^{-1}\left( D^2 - R^2 \right)\\i\alpha \mathcal{U} - Re^{-1} D^2 \end{matrix}}
\right.\hspace{-10.3em}
\begin{matrix}
  i\alpha \mathcal{U} - Re^{-1}\left( D^2 - R^2 \right) & 2 Re^{-1}R^2 D^{(1)}_\theta & 0 \\
  -2 Re^{-1}R^2 D^{(1)}_\theta & i\alpha \mathcal{U} - Re^{-1}\left( D^2 - R^2 \right) & 0 \\
  \left( D^{(1)}_r \mathcal{U} \right) & \left( D^{(1)}_\theta \mathcal{U} \right) & i\alpha \mathcal{U} - Re^{-1} D^2
\end{matrix}
\hspace{-10.3em}
\left.\phantom{\begin{matrix}i\alpha \mathcal{U} - Re^{-1}\left( D^2 - R^2 \right) \\i\alpha \mathcal{U} - Re^{-1}\left( D^2 - R^2 \right)\\i\alpha \mathcal{U} - Re^{-1} D^2 \end{matrix}}\right]\hspace{-1em}
\begin{tabular}{l}
  $\scriptstyle \phantom{\begin{matrix} i\alpha \mathcal{U} - Re^{-1}\left( D^2 - R^2 \right) \ \end{matrix}} \hspace{-10em}\big\rbrace N$\\
  $\scriptstyle \phantom{\begin{matrix} i\alpha \mathcal{U} - Re^{-1}\left( D^2 - R^2 \right)\ \end{matrix}} \hspace{-10em}\big\rbrace N$\\
  $\scriptstyle \phantom{\begin{matrix} i\alpha \mathcal{U} - Re^{-1} D^2 \ \end{matrix}} \hspace{-6.5em}\big\rbrace N$
\end{tabular}\\
\begin{tabular}{lll}
  \hspace{-30.5em}\vspace{-5em}
  $\underbrace{\phantom{i\alpha \mathcal{U} - Re^{-1}\left( D^2 - R^2 \right)}}_{N}$ & $\underbrace{\phantom{i\alpha \mathcal{U} - Re^{-1}\left( D^2 - R^2 \right)}}_{N}$ & $\underbrace{\phantom{\left( i\alpha \mathcal{U} - Re^{-1} D^2\right)}}_{N}$,
\end{tabular}
\]\vspace{1em}

\noindent with $D^2=D^{(2)}_r + RD^{(1)}_r + R^2 D^{(2)}_\theta - \alpha^2 I$
and $\mathcal{U} = \text{diag}\left( U_j \right)$ for $j=0,1,\dots N-1$.
The matrices appearing in (\ref{eq:disc-pert-u-p}) are then assembled with the
boundary condition for $\hat{\bm{u}}$ taken into account and read
\[
S'=\left[\phantom{\begin{matrix} S'_{11} \\ I \\ S'_{11} \\ I \\ S'_{11} \\ I \end{matrix}}
\right.\hspace{-1.7em}
\begin{matrix}
  S'_{11} & S'_{12} & S'_{13} \\
  I & 0 & 0 \\
  S'_{21} & S'_{22} & S'_{23} \\
  0 & I & 0 \\
  S'_{31} & S'_{32} & S'_{33} \\
  0 & 0 & I
\end{matrix}
\hspace{-1.7em}
\left.\phantom{\begin{matrix} S'_{11} \\ I \\ S'_{11} \\ I \\ S'_{11} \\ I \end{matrix}}\right]\hspace{-1em}
\begin{tabular}{l}
  $\scriptstyle \phantom{\begin{matrix} S'_{11} \ \end{matrix}} \hspace{-1.5em}{\displaystyle\rbrace} \left( N_r - 1 \right) N_\theta$\\
  $\scriptstyle \phantom{\begin{matrix} I \ \end{matrix}} \hspace{-0.55em}{\displaystyle\rbrace} N_\theta$\\
  $\scriptstyle \phantom{\begin{matrix} S'_{11} \ \end{matrix}} \hspace{-1.5em}{\displaystyle\rbrace} \left( N_r - 1 \right) N_\theta$\\
  $\scriptstyle \phantom{\begin{matrix} I \ \end{matrix}} \hspace{-0.55em}{\displaystyle\rbrace} N_\theta$\\
  $\scriptstyle \phantom{\begin{matrix} S'_{11} \ \end{matrix}} \hspace{-1.5em}{\displaystyle\rbrace} \left( N_r - 1 \right) N_\theta$\\
  $\scriptstyle \phantom{\begin{matrix} I \ \end{matrix}} \hspace{-0.55em}{\displaystyle\rbrace} N_\theta$
\end{tabular},\\
\hspace{2em}
T=\left[\phantom{\begin{matrix} S'_{11} \\ I \\ S'_{11} \\ I \\ S'_{11} \\ I \end{matrix}}
\right.\hspace{-1.7em}
\begin{matrix}
  D^{(1)}_r \\ 0 \\ R D^{(1)}_\theta \\ 0 \\ i\alpha I \\ 0
\end{matrix}
\hspace{-1.7em}
\left.\phantom{\begin{matrix} S'_{11} \\ I \\ S'_{11} \\ I \\ S'_{11} \\ I \end{matrix}}\right],\hspace{-1em}
\hspace{2em}
F=\left[\phantom{\begin{matrix} D^{(1)}_\theta \end{matrix}}
\right.\hspace{-1.7em}
\begin{matrix}
  D^{(1)}_r + R & R D^{(1)}_\theta & i\alpha I
\end{matrix}
\hspace{-1.7em}
\left.\phantom{\begin{matrix} D^{(1)}_\theta \end{matrix}}\right]\hspace{-1em}
\begin{tabular}{l}
  $\scriptstyle \phantom{\begin{matrix} D^{(1)}_r + R \ \end{matrix}} \hspace{-3.8em}\big\rbrace N$
\end{tabular}\\
\begin{tabular}{llll}
  \hspace{-36.1em}\vspace{-6.9em}
  $\underbrace{\phantom{S'_{11}}}_{N}$\hspace{-0.1em} & $\underbrace{\phantom{S'_{11}}}_{N}$\hspace{-0.1em} & $\underbrace{\phantom{S'_{11}}}_{N}$\hspace{9.4em} & $\underbrace{\phantom{R D^{(1)}_\theta}}_{N}$
\end{tabular}
\begin{tabular}{lll}
  \hspace{-12.2em}\vspace{-1.8em}
  $\underbrace{\phantom{D^{(1)}_r + R}}_{N}\,$ & $\underbrace{\phantom{R D^{(1)}_\theta}}_{N}\,$ & $\underbrace{\phantom{D^{(1)}_\theta}}_{N}$
\end{tabular}
\hspace{0em}.
\]
\vspace{0.5em}

In total, the system (\ref{eq:disc-pert-u-p}) contains $4N$ unknowns and the
boundary conditions are treated as additional linear equations. To eliminate the pressure from the system and reduce the memory requirements, we use the method of algebraic
reduction described in \cite{Boiko2008}. To that end we perform the QR decomposition of
matrices $T$ and $F^*$ as (here * denotes conjugate transpose):
\[
T = T_Q T_R = \left[\phantom{\begin{matrix} T_{Q,1} \end{matrix}}
\right.\hspace{-2.2em}
\begin{matrix}
  T_{Q,1} & T_{Q,2}
\end{matrix}
\hspace{-2.2em}
\left.\phantom{\begin{matrix} T_{Q,1} \end{matrix}}\right]\hspace{0em}
\left[\phantom{\begin{matrix} T_{R,1} \\ 0 \end{matrix}}
\right.\hspace{-2.2em}
\begin{matrix}
  T_{R,1} \\ 0
\end{matrix}
\hspace{-2.2em}
\left.\phantom{\begin{matrix} T_{R,1} \\ 0 \end{matrix}}\right]
\begin{tabular}{l}
  $\scriptstyle \phantom{\begin{matrix} T_{R,1} \ \end{matrix}} \hspace{-2.8em}\big\rbrace N$\\
  $\scriptstyle \phantom{\begin{matrix} 0 \ \end{matrix}} \hspace{-1.4em}\big\rbrace 2N$\\
\end{tabular}\\
\begin{tabular}{ll}
  \hspace{-9.9em}\vspace{-1.5em}
  $\underbrace{\phantom{T_{Q,1}}}_{N}$ & $\underbrace{\phantom{T_{Q,2}}}_{2N}$
\end{tabular}
\begin{tabular}{l}
  \hspace{-4.6em}\vspace{-2.7em}
  $\underbrace{\phantom{T_{Q,1}}}_{N}$
\end{tabular},\quad
F^* = F_Q F_R = \left[\phantom{\begin{matrix} F_{Q,1} \end{matrix}}
\right.\hspace{-2.2em}
\begin{matrix}
  F_{Q,1} & F_{Q,2}
\end{matrix}
\hspace{-2.2em}
\left.\phantom{\begin{matrix} F_{Q,1} \end{matrix}}\right]\hspace{0em}
\left[\phantom{\begin{matrix} F_{R,1} \\ 0 \end{matrix}}
\right.\hspace{-2.2em}
\begin{matrix}
  F_{R,1} \\ 0
\end{matrix}
\hspace{-2.2em}
\left.\phantom{\begin{matrix} T_{R,1} \\ 0 \end{matrix}}\right]
\begin{tabular}{l}
  $\scriptstyle \phantom{\begin{matrix} F_{R,1} \ \end{matrix}} \hspace{-2.8em}\big\rbrace N$\\
  $\scriptstyle \phantom{\begin{matrix} 0 \ \end{matrix}} \hspace{-1.4em}\big\rbrace 2N$\\
\end{tabular}\\
\begin{tabular}{ll}
  \hspace{-10.2em}\vspace{-1.5em}
  $\underbrace{\phantom{F_{Q,1}}}_{N}$ & $\underbrace{\phantom{F_{Q,2}}}_{2N}$
\end{tabular}
\begin{tabular}{l}
  \hspace{-4.7em}\vspace{-2.7em}
  $\underbrace{\phantom{F_{Q,1}}}_{N}$
\end{tabular},
\]
\vspace{1em}

\noindent where $T_Q$ and $F_Q$ are unitary matrices of size $3N$ with orthonormal
columns and $T_{R,1}$ and $F_{R,1}$ are non-singular upper triangular
matrices of size $N$. An alternative set of $2N$ unknowns $\hat{\bm{W}}$ can then be defined using the transformations
\begin{equation}
  \hat{\bm{W}} = F_{Q,2}^* \hat{\bm{U}} \quad \text{and} \quad \hat{\bm{U}} = F_{Q,2} \hat{\bm{W}}.
\end{equation}
Upon substituting the last expression into the Eq.~(\ref{eq:disc-ns}) and making use of the QR decomposition of $T$, we
obtain
\begin{equation}
  \left( i\alpha c - S' \right) F_{Q,2} \hat{\bm{W}} - T_{Q,1}T_{R,1}\hat{P} = 0.
\end{equation}
Multiplying this matrix equation with $T_{Q,2}^*$ yields,
%\begin{equation}
%  \hat{P} = T_{R,1}^{-1}\left[ T_{Q,1}^* \left( i\alpha c - S' \right) F_{Q,2}\hat{\bm{W}} \right]
%\end{equation}
%and
\begin{equation}
  T_{Q,2}^* \left( i\alpha c - S' \right) F_{Q,2} \hat{\bm{W}} = 0,
\end{equation}
where we have used the orthonormality of the columns of $T_Q$.
With similar notations as in \cite{Boiko2008}, we rewrite
the above equation as follows:

\begin{eqnarray}
    i\alpha c \hat{\bm{W}} = H \hat{\bm{W}},\label{eq:disc-pert-w-1}%\\
    %\hat{P} = T_{R,1}^{-1}\left[ \left( i\alpha c A_{12} - S'_{12} \right) \hat{\bm{W}} \right],
\end{eqnarray}
where $H = A_{22}^{-1} M_{22}$, $A_{22}=T_{Q,2}^* F_{Q,2}$ and
$M_{22}=T_{Q,2}^* S' F_{Q,2}$ (the non-singularity of $A_{22}$ is
a consequence of the non-singularity of the matrix $FT$ as shown in \cite{Boiko2008}).
Solving this eigenvalue problem of size $2N$ significantly reduces the
computational effort and by construction only yields the finite part of the spectrum of the original system (\ref{eq:disc-pert-u-p}).
The discrete operators appearing in Eq.~(\ref{eq:disc-pert-w-1}) are implemented in an in-house Python solver relying on the NumPy library \cite{NumPy2020}.
Depending on the convergence properties observed for different flow regimes, we then solve the complex
matrix eigenvalue problem (\ref{eq:disc-pert-w-1})
using the packages SciPy \cite{SciPy2020} or SLEPc \cite{SLEPc2005}.

\begin{figure}[ht!]
  \includegraphics[width=0.6\textwidth]{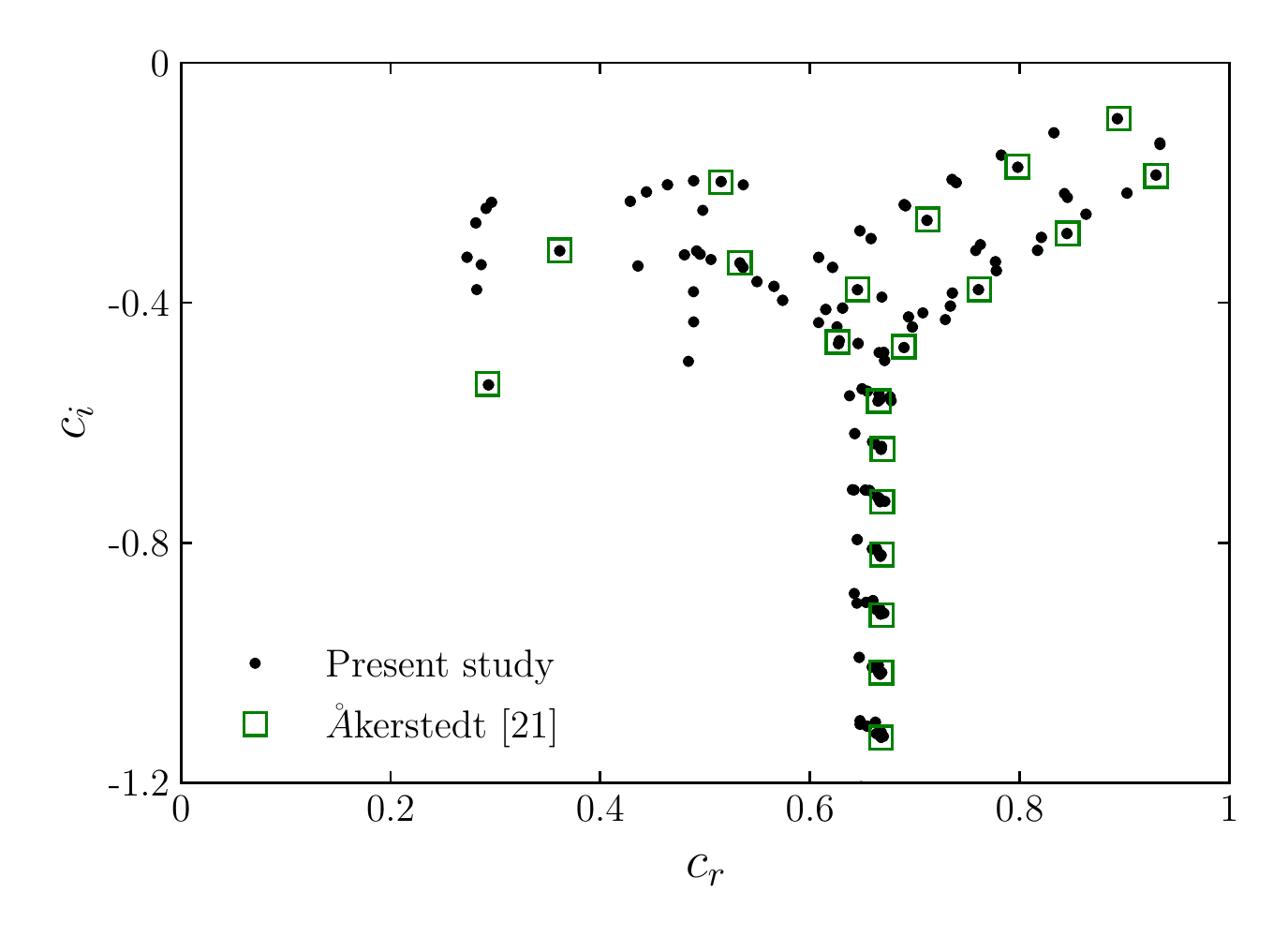}
  \caption{Eigenvalue spectra for the case of the MHD pipe flow with
  uniform axial magnetic field for $Re=1000$, $\alpha=1$, $Ha= 44.7$
  and $\chi=\infty$. Results obtained in this paper are compared
  with those reported in \cite{Akerstedt1995} for azimuthal wavenumber $n=1$.}
  \label{fig:aker-bench}
\end{figure}

The numerical procedure has been validated in two stages. First, we compared our solution of
the base flow equations with existing numerical \cite{Vantieghem2009} and analytical
solutions \cite{Gold1962, Ihara1967}. Second, we tested our formulation of the linear
stability problem using the results of the linear stability analysis of Hagen-Poiseuille
flow \cite{Schmid1994,Meseguer2003}, as well as the MHD pipe flow subject
to an axial magnetic field \cite{Akerstedt1995}.
In Table~\ref{tab:convergence}
we compare our results and those of Schmid \& Henningson \cite{Schmid1994}
for the linear stability analysis of Hagen-Poiseuille flow.
We observe that we reach an agreement with 9 significant digits in the decay
rates with a maximum of 26 Chebyshev collocation points in the
computational domain $r > 0$.
\begin{table}
  \caption{\label{tab:convergence}Convergence of the least stable eigenvalues in the case of
  Hagen-Poiseuille flow for $Re=3000$ and $\alpha=1$. $N_r$ is the number of
  Chebyshev collocation points in the computational domain. The result obtained in
  this paper using a 2D spectral code are compared with those reported
  in \cite{Schmid1994}.}
  \begin{ruledtabular}
    \begin{tabular}{ccc}
      Schmid \& Henningson \cite{Schmid1994} & \multicolumn{2}{c}{Present study}\\
      \cline{1-1} \cline{2-3}
      $c$, $n=0$&$N_r$&$c$\\
      \hline
      $(0.94836022,-0.05197311)$ & 15 & $(0.94851426,-0.05185138)$\\
      & 19 & $(0.94836026,-0.05197305)$\\
      & 23 & $(0.94836022,-0.05197311)$\\
      \hline
      $c$, $n=1$&&\\
      \hline
      $(0.91146557,-0.04127564)$ & 18 & $(0.91148503,-0.04126378)$\\
      & 22 & $(0.91146537,-0.04127591)$\\
      & 26 & $(0.91146557,-0.04127564)$\\
      \hline
      $c$, $n=2$&&\\
      \hline
      $(0.88829766,-0.06028569)$ & 16 & $(0.88829953,-0.06032205)$\\
      & 20 & $(0.88829742,-0.06028591)$\\
      & 24 & $(0.88829766,-0.06028569)$\\
    \end{tabular}
  \end{ruledtabular}
\end{table}
The eigenvalue spectrum for the case of the MHD pipe flow subject to a
uniform axial magnetic field $\bm{1}_{B_0} = \bm{1}_z$ is shown in
Fig.~\ref{fig:aker-bench}. Our results obtained using a two-dimensional formulation of the MHD equations are compared
with those reported in \cite{Akerstedt1995} for an azimuthal wavenumber $n=1$.
Unlike the MHD pipe flow with a transverse magnetic field, in this case,
the base flow is not modified by the magnetic field. Thus, the problem
is characterized by homogeneous axial and azimuthal directions. We
see that all the eigenvalues found by {\AA}kerstedt \cite{Akerstedt1995} for $n=1$
using the one-dimensional formulation appear in the full spectrum that also captures all the other possible azimuthal wavenumbers.
Throughout this work we have naturally observed that the number of grid points necessary to converge the results presented increase with Reynolds and Hartmann numbers. This is particularly true in the conducting case characterized by the gradual emergence of
zones of velocity overspeed in the Roberts layers. Typically, to converge the reported growth rates with 7
significant digits, we used $N_\theta$ and $N_r$ in the ranges $40$ to $120$ and  $35$ to $60$, respectively.

\section{Results}\label{section:results}
\subsection{Base flow profile}
Unlike the MHD pipe flow subject to a streamwise magnetic
field studied by {\AA}kerstedt \cite{Akerstedt1995}, the base flow velocity
profile no longer possesses axial symmetry when a transverse magnetic field is applied. In general, it becomes elongated in the direction of the magnetic field and may also be characterized by
the presence of the regions of velocity overspeed when the walls have non-zero conductivity.
The latter effect occurs for certain flow parameters when the electric current enters
the pipe's wall and leads to higher flow
velocities in the side layers (also known as Roberts layers) than in the core region \cite{Samad1981, Vantieghem2009}.

\begin{figure}[ht!]
  \begin{tikzpicture}
    \node[anchor=south west,inner sep=0] at (0,0) {\includegraphics[width=0.95\textwidth]{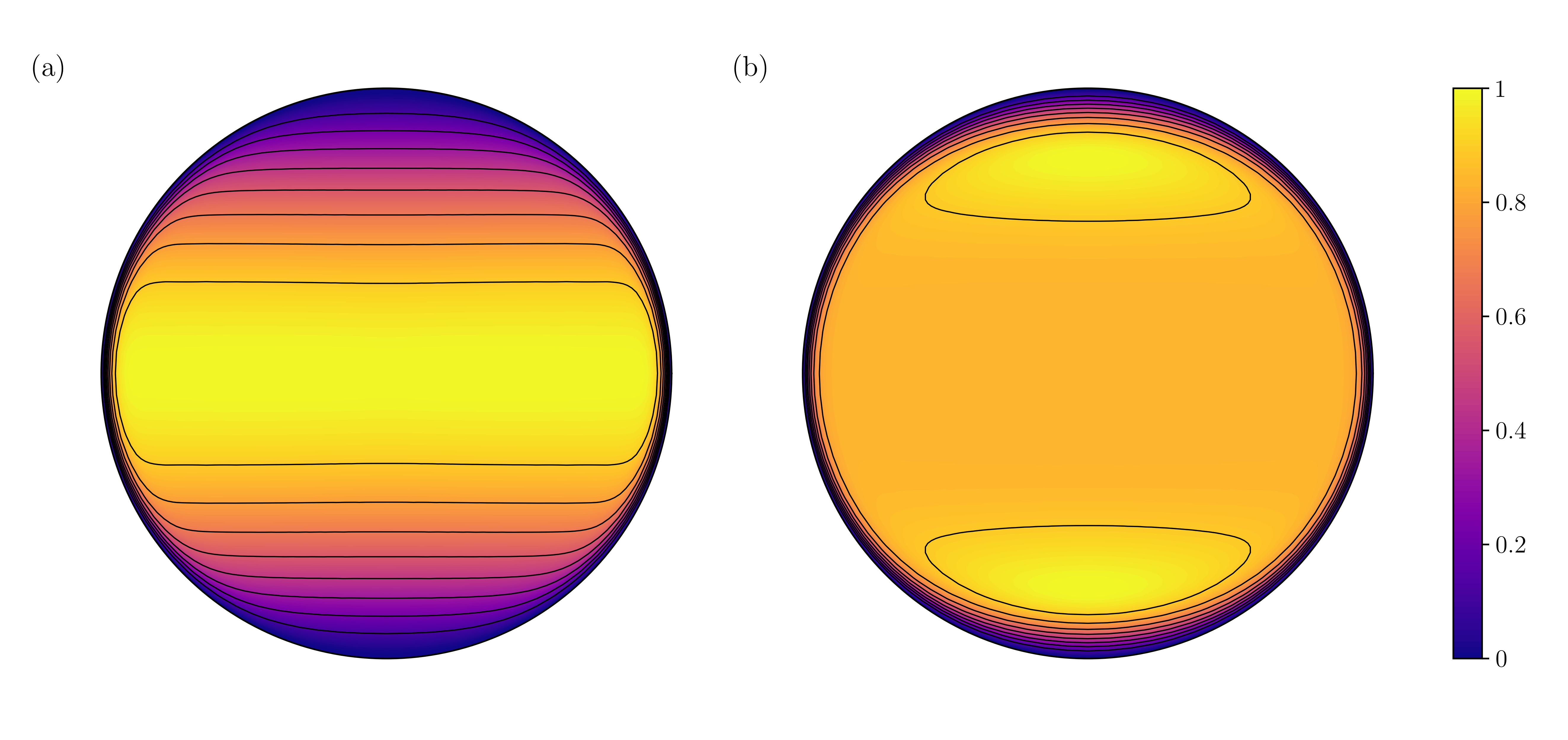}};
    \draw[line width=1pt, -stealth](7,0.5)-- +(2,0) node[pos=0.5, above] {$\bm{B}_0$};
  \end{tikzpicture}
  \caption{Contours of the base flow velocity $U(r,\theta)$ for $Ha=60$ and
  (a) $\chi = 0$ and (b) $\chi = \infty$.}
  \label{fig:base-flow-contours}
\end{figure}

\begin{figure}[ht!]
  \includegraphics[width=0.85\textwidth]{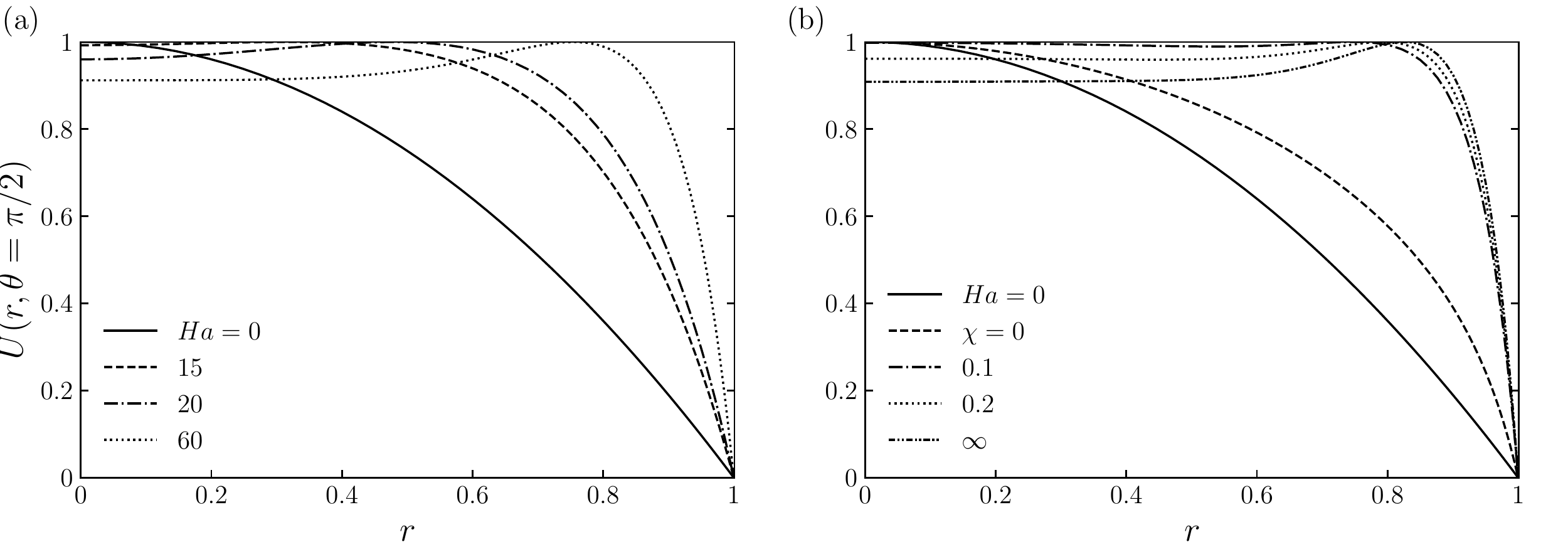}
  \caption{Emergence of zones of velocity overspeed in the base velocity
  profile (a) with increasing $Ha$ at $\chi=\infty$; (b) with increasing
  $\chi$ at $Ha=100$.}
  \label{fig:base-flow}
\end{figure}

To illustrate this important feature, we show in Fig.~\ref{fig:base-flow-contours} the
contours of the base flow velocity $U(r,\theta)$ at
$Ha=60$ for $\chi=0$ and $\chi = \infty$. In both cases the profiles are elongated in the direction of the magnetic field, but overspeeds in the side layers only appear when the wall is conducting. The detailed criteria for the occurrence of velocity overspeeds may be found in Vantieghem {\em et al.} \cite{Vantieghem2009}. In particular these authors have shown that the overspeed regions emerge only for $Ha > 12$ and for non-zero conductivity. For $Ha \geq 250$, the minimal conductivity needed to observe them scales as $Ha^{-2/3}$. The requirement of minimal Hartmann number and finite conductivity for the existence of velocity overspeeds is further demonstrated in Figs.~\ref{fig:base-flow}(a) and \ref{fig:base-flow}(b), where we show their gradual formation with increasing $Ha$ and $\chi$, respectively.

\subsection{Mechanism of instability}
\begin{figure}[ht!]
  \includegraphics[width=0.85\textwidth]{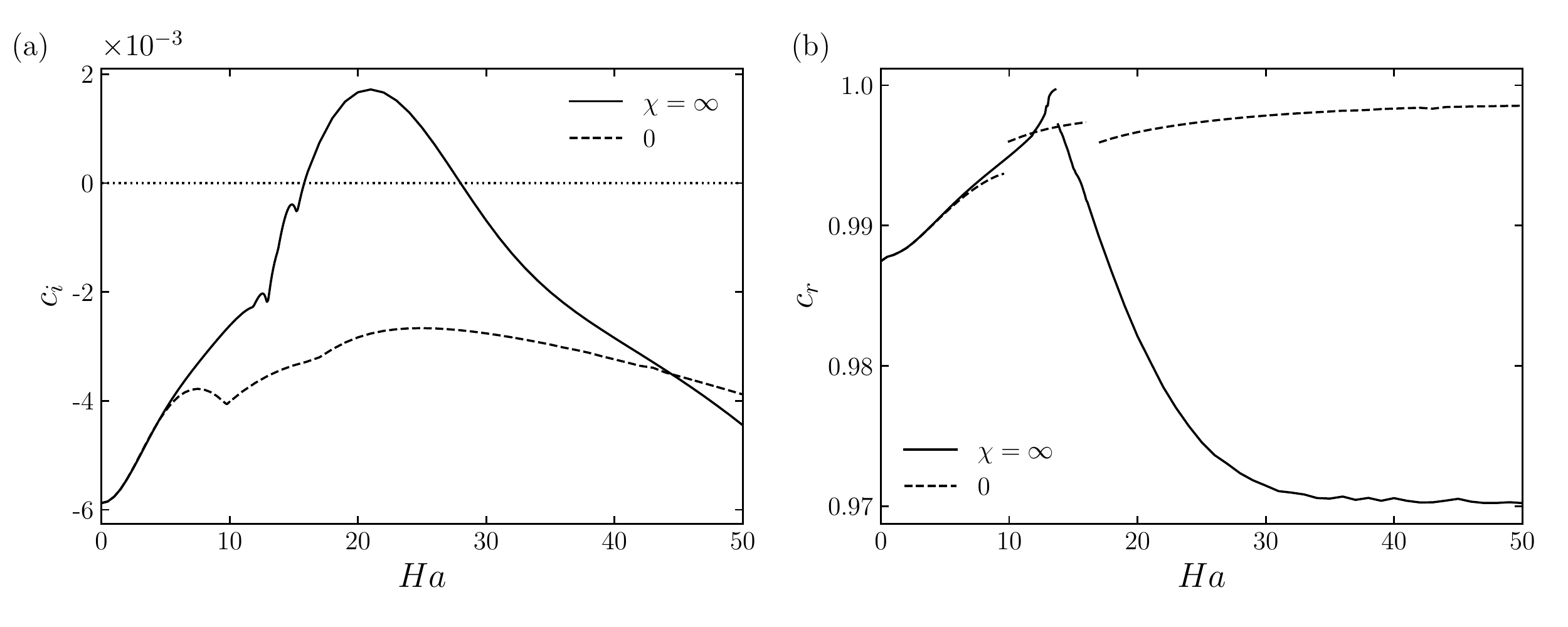}
  \caption{(a) Growth rate $c_i$ and (b) phase velocity $c_r$ of the leading
  eigenmode as functions of the Hartmann number $Ha$ for $\chi=0$ (dashed curve)
  and $\chi=\infty$ (solid curve) at $Re=10^5$ and $\alpha=1.5$.}
  \label{fig:ci-vs-ha}
\end{figure}

The stability of the flow is determined by the global eigenvalue $c$ with
the largest imaginary part $c_i$. In Fig.~\ref{fig:ci-vs-ha}(a) we plot the maximum
value of $c_i$ as a function of the Hartmann number $Ha$ for $Re=10^5$ and $\alpha=1.5$
in the cases of perfectly insulating and perfectly conducting walls.
In the absence of a magnetic field, Hagen-Poiseuille flow is linearly
stable with $\max (c_i)=-5.874 \cdot 10^{-3}$. From the figure we observe that for both conductivities the decay rate of the leading
eigenmode in the MHD case exceeds this value on the interval of moderate
Hartmann numbers $0 < Ha \le 50$. This indicates that due to mean flow deformations, the MHD pipe flow with a transverse magnetic field is
less stable than its hydrodynamic counterpart. For
small values of $Ha$, the curves corresponding to the two cases are indistinguishable. In this
regime, the modification of the base velocity profile by the magnetic
field mainly consists in its elongation along $\bm{B}_0$ for all values of
$\chi$. For $Ha \gtrsim 5$, we see that the destabilizing
effect of the magnetic field becomes more pronounced for $\chi = \infty$. In that case, the flow even becomes linearly unstable for $16 \le Ha \le 23$ whereas it remains stable for all values of $Ha$ when $\chi =0$.

We associate the instability of the flow in the conducting case with the emergence of the overspeed regions and the inflection points present in the velocity profile. First, the instability occurs just after the overspeed regions form and these exist only in the conducting case. Second, looking at Fig.~\ref{fig:ci-vs-ha}(b) we observe that the phase velocity of the most unstable mode in the conducting case is very similar to that of the flow around the inflection point in the velocity profile (see Fig. \ref{fig:base-flow}(a)). We also note that the leveling of $c_r$ for $Ha\gtrsim 30$ is consistent with the results of Vantieghem {\em et al.} who showed that for high Hartmann numbers, the ratio of the velocity in the overspeed regions to that in the core region converges to a fixed value.

For MHD flows, the stability properties depend on the relative roles of mean flow deformations and the magnitude of Joule damping. To isolate the impact of mean flow deformations, we have conducted a numerical experiment in which we have suppressed the electromagnetic term in Eq.~(\ref{eq:pert-2}) and solved the hydrodynamic stability problem with the base flow computed from system (\ref{eq:base-flow}). Some results are illustrated in Table~\ref{tab:num-experiment}
for $Re = 10^4$, $\alpha = 1.5$, $\chi = 0, \infty$ and different values
of $Ha$. In the perfectly insulating case, increasing the Hartmann number leads to the thinning of the Hartmann
layers and the destabilization of the flow. However, the decay rate
of the leading eigenvalue becomes nearly invariant for large
values of $Ha$. Therefore, even in the absence of Joule damping, the base flow modifications in the insulating case -- and the absence of overspeed regions -- are not making the flow unstable. Note that this conclusion has been tested for Reynolds numbers up to $10^6$.
In the perfectly conducting case, suppressing the
Joule dissipation results as expected in an increase of the growth rate of the leading eigenmode and to the possibility of instability at lower Reynolds numbers.

\begin{table}
  \caption{\label{tab:num-experiment} Maximum eigenvalues with and without the effect of magnetic damping for different values of $Ha$ in the cases of perfectly insulating and perfectly conducting walls.}
\begin{ruledtabular}
  \begin{tabular}{cccc}
    $\chi$ & $Ha$ & $c_i$ with magnetic damping & $c_i$ without magnetic damping \\
    \hline
    0 & 15 & -0.015072 & -0.007913\\
    & 60 & -0.020923 & -0.004959\\
    & 200 & -0.030967 & -0.004792\\
    $\infty$ & 15 & -0.00705 & -0.002428\\
    & 60 & -0.015753 & 0.003715
  \end{tabular}
\end{ruledtabular}
\end{table}

The influence of Joule damping is also clear when looking again at Fig.~\ref{fig:ci-vs-ha}(a), in which we see that the dependence of $c_i$ on
$Ha$ is non-monotonic. On the interval
$0 \le Ha \le 50$, the global maxima occur for $Ha=25$ and $Ha=21$ in the
perfectly insulating and perfectly conducting cases, respectively. Beyond those values, the destabilization of the flow due to the modifications of the base velocity profile is suppressed by Joule damping, which has an opposite effect on the stability of the flow. Notably, in the perfectly conducting case, the stabilization effect of the magnetic damping becomes dominant for $Ha\gtrsim 30$ when the overspeed regions have reached their maximum amplitude with respect to the core flow.

\begin{figure}[ht!]
  \includegraphics[width=0.9\textwidth]{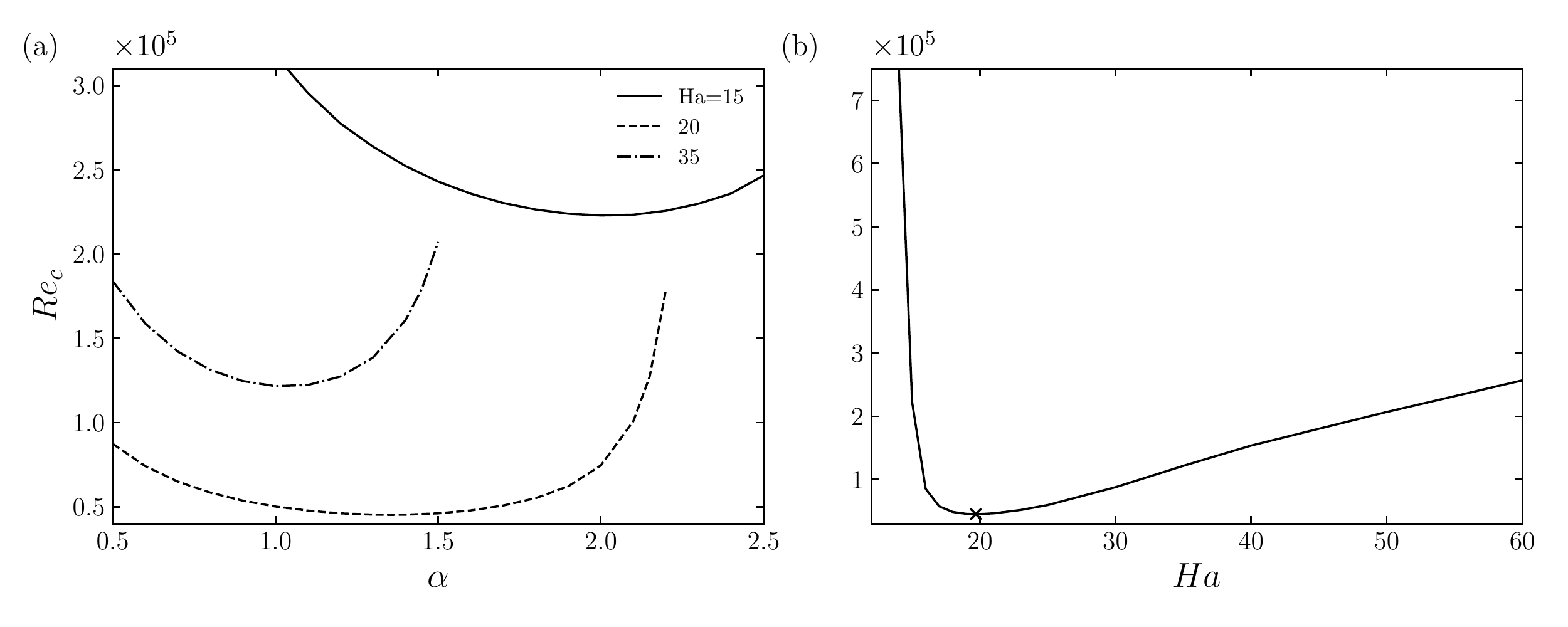}
  \caption{(a) Marginal Reynolds number $Re$ as a function of the axial
  wavenumber $\alpha$ for $\chi=\infty$ and different values of $Ha$. (b) Critical Reynolds number $Re_c$ optimized over all $\alpha$ as a function of the Hartmann number $Ha$ for $\chi=\infty$. The marker $\times$ corresponds to the critical Reynolds number $Re_c=45230$ occurring for $Ha=19.7$ and $\alpha=1.37$.}
  \label{fig:neutral-curves}
\end{figure}

To conclude this section, we plot in Fig.~\ref{fig:neutral-curves} the marginal stability curves for $\chi=\infty$. For a given
value of the axial wavenumber $\alpha$, the marginal state is determined by the critical Reynolds number $Re_c$ for which the growth rate $c_i = 0$.
$Re_c$ as a function of $\alpha$ for different values of $Ha$ is shown in Fig.~\ref{fig:neutral-curves}(a).
For $Ha=15$, the marginal Reynolds numbers are of order of magnitude $10^5$ for $1 \le \alpha \le 2.5$. As detailed above,
an increase in the Hartmann number first makes the flow more unstable while at
higher Hartmann numbers the flow is stabilized due to stronger magnetic damping. In Fig.~\ref{fig:neutral-curves}(b), we illustrate $Re_c$ minimized over all $\alpha$ as a function of $Ha$ for moderate values of $Ha$. The minimum of this curve corresponds to
the global critical Reynolds number $Re_c=45230$ occurring for $Ha=19.7$ and $\alpha = 1.37$.
\subsection{Linear stability for finite $\chi$}
\label{section:finite-chi}
\begin{figure}[ht!]
  \includegraphics[width=0.6\textwidth]{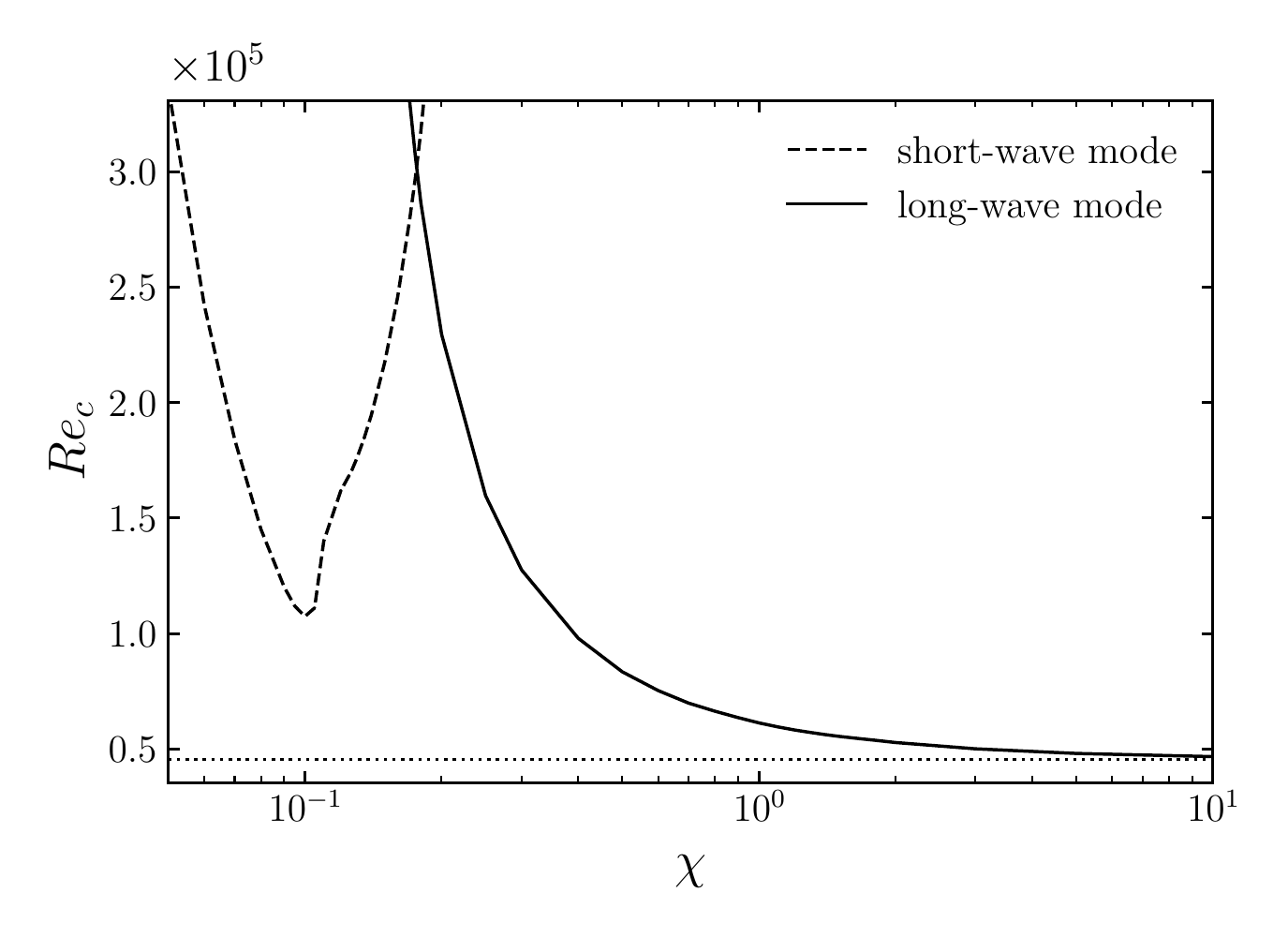}
  \caption{Critical Reynolds number $Re_{c}$ as a function of wall
  conductance ratio $\chi$. The dashed and the solid curves correspond
  to the two regimes of instability. The dotted line marks the global critical Reynolds number
  $Re_{c} = 45230$ occurring for $\chi=\infty$.}
  \label{fig:re-vs-chi}
\end{figure}

In this section, we analyze the influence of the wall conductance ratio $\chi$ on the stability characteristics of the MHD pipe flow. Depending on $\chi$, two different regimes can be identified in which the most unstable mode is associated respectively with larger
or smaller values of the axial wavenumber $\alpha$ (and hence shorter and longer axial
wavelengths).
%
%
%, higher and lower Hartmann numbers.
In Fig.~\ref{fig:re-vs-chi}, we plot the corresponding global critical Reynolds
number minimized over all values of $Ha$ and $\alpha$. The short-wave mode regime is the most
unstable for $0 < \chi < 0.176$. The corresponding $Re_{c}$ curve attains a
minimum value of 107420 at $\chi = 0.1$. For $\chi > 0.176$,
the stability threshold is defined by the long-wave mode regime, and with the increase of
$\chi$, the critical Reynolds number converges rapidly towards the global minimum value
of $Re_c=45230$ associated with $\chi=\infty$.

\begin{figure}[ht!]
  \includegraphics[width=0.8\textwidth]{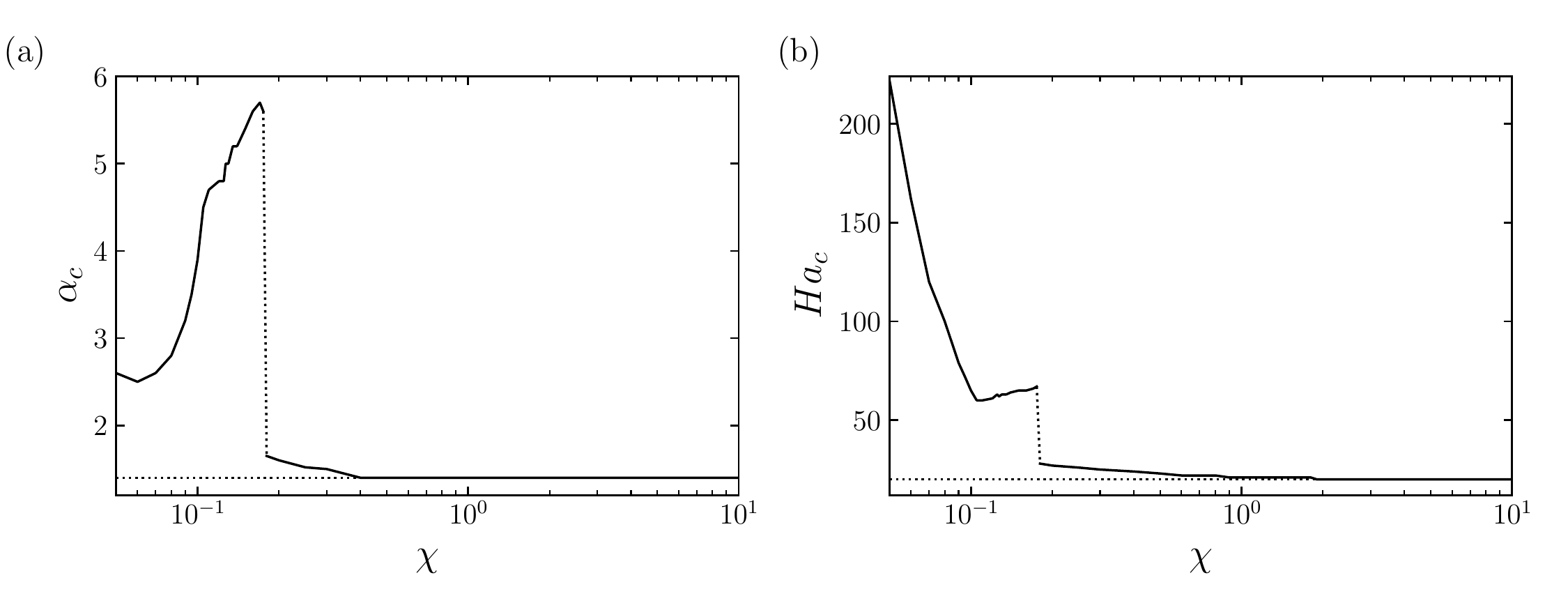}
  \caption{(a) Critical axial wavenumber $\alpha_c$ and (b) critical Hartmann
  number $Ha_c$ as functions of wall conductance ratio $\chi$. The discontinuities
  in the curves occur due to the change in the regime of instability.}
  \label{fig:alpha-ha-vs-chi}
\end{figure}

The critical wavenumber $\alpha_{c}$ and the critical Hartmann number $Ha_{c}$
corresponding to the most unstable mode at different values of $\chi$ are shown
in Figs.~\ref{fig:alpha-ha-vs-chi}(a) and \ref{fig:alpha-ha-vs-chi}(b), respectively. In both plots, the discontinuity occurring
for $\chi=0.176$ marks the transition between the short- and long-wave regimes.
Lower values of wall conductance ratio and short-wave modes are therefore
associated with higher values of $Ha$. This behavior is consistent with the results
of Vantieghem \textit{et al.} \cite{Vantieghem2009} who showed that the Hartmann number marking the
emergence velocity overspeeds in the base velocity profile
increases when $\chi$ decreases (except around a local minimum at $Ha=35$) \cite{Vantieghem2009}. At large values
of $\chi$, i.e, in the long-wave regime, $\alpha_{c}\approx 1.4$
and $Ha_{c} \approx 20$ for $\chi\gtrsim 0.4$ and $\chi \gtrsim 1.9$, respectively.

\subsection{Characteristics of the most unstable perturbation}
\begin{figure}[ht!]
  \begin{tikzpicture}
    \node[anchor=south west,inner sep=0] at (0,0) {\includegraphics[width=\textwidth]{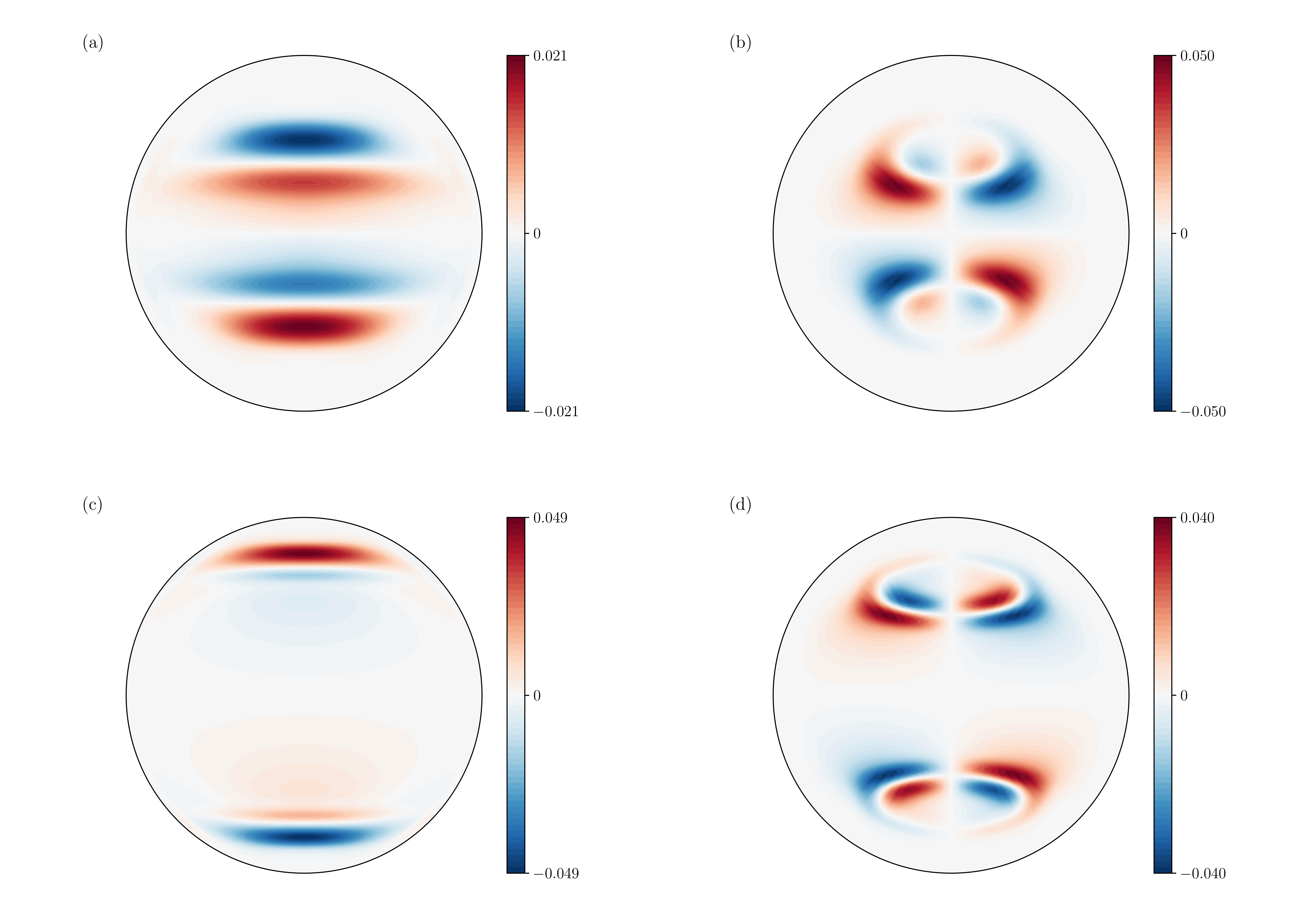}};
    \draw[line width=1pt, -stealth](8,7.5)-- +(2,0) node[pos=0.5, above] {$\bm{B}_0$};
  \end{tikzpicture}
  \caption{Contours of the streamwise velocity of the marginal mode
  at $z=0$ for (a) $\chi=0.1$, $Ha=65$; (b) $\chi=\infty$,
  $Ha=20$; (c) $\chi=0.05$, $Ha=222$;
  (d) $\chi=\infty$, $Ha=35$.}
  \label{fig:streamwise-velocity-contours}
\end{figure}

We now discuss the structure of the most unstable perturbations. When $Ha>0$, the perturbation
equations (\ref{eq:pert})
admit solutions of four different symmetry types, analogous to those described
by Tatsumi \cite{Tatsumi1990} for the case of the flow in a duct of square or rectangular
cross-section. In Figs.~\ref{fig:streamwise-velocity-contours}(a) and \ref{fig:streamwise-velocity-contours}(b) we plot respectively
the contours of the streamwise velocity perturbation of the most unstable
short- and long-wave modes occurring for $\chi=0.1$ and $\chi=\infty$. We note that they possess different types of axial symmetry. The streamwise
velocity of a short-wave mode is antisymmetric and symmetric (A-S) with respect to
the Cartesian axes $x$ and $y$, respectively, whereas that of a long-wave
mode is antisymmetric (A-A) with respect to both.
In Tatsumi \cite{Tatsumi1990} and later in Priede {\em et al.} \cite{Priede2010} in the
context of the MHD duct flow, these A-S and A-A modes were labeled modes I and II,
respectively.
The above observation is systematic, and we always find that in the short-wave regime of instability, the most unstable mode is symmetric along the magnetic field, whereas in the long-wave regime, the stability threshold
is defined by the A-A mode with opposite symmetry with respect to the $y$-axis.
This property is consistent with the fact that the long-wave mode regime prevails only for small Hartmann numbers and large values of $\chi$ (see section \ref{section:finite-chi}). As a consequence, given their A-A symmetry, such modes experience high Joule damping and are rapidly damped by strong magnetic fields.

From Figs.~\ref{fig:streamwise-velocity-contours}(c) and \ref{fig:streamwise-velocity-contours}(d) we further see how the
velocity distribution of the most unstable perturbation changes compared to that
in Figs.~\ref{fig:streamwise-velocity-contours}(a) and \ref{fig:streamwise-velocity-contours}(b) as we increase the Hartmann
number. Since the perturbations are localized in the regions of velocity
overspeed, they are progressively expelled from the core of the flow as the
Roberts layers get thinner with an increasing magnetic field.

\begin{figure}[ht!]
  \includegraphics[width=0.6\textwidth]{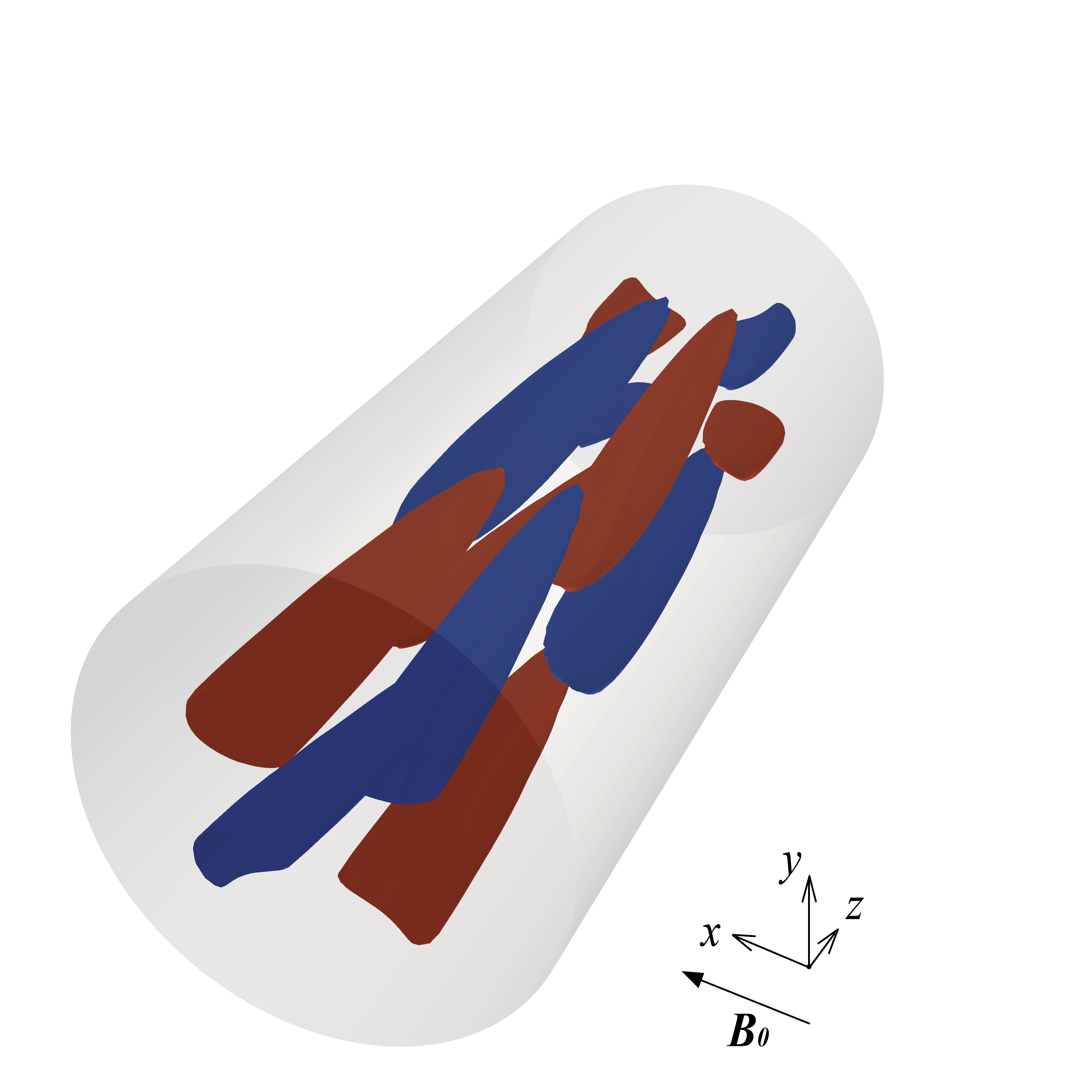}
  \caption{Isosurfaces of the streamwise velocity component of the most
  unstable long-wave perturbation, consisting in low- (blue) and high- (red) speed streaks. The perturbation is shown for one axial wavelength, at
  $\chi=\infty$, $Ha=20$, $\alpha=1.4$ and $Re_c=45400$.}
  \label{fig:3D-view-long-wave}
\end{figure}

The energy partition among the three velocity components also differs
significantly when considering short- and long-wave critical modes. Most of the
long-wave mode kinetic energy, averaged over an axial wavelength, is
concentrated in the streamwise velocity component. The perturbations mainly
consist of pairs of high- and low-speed longitudinal streaks as shown in
Fig.~\ref{fig:3D-view-long-wave}. A detailed exam of the flow streamlines also
shows that in regions of low streamwise velocity, the other two velocity
components allow the transfer of fluid particles from one streak to another.

\begin{figure}[ht!]
  \includegraphics[width=\textwidth]{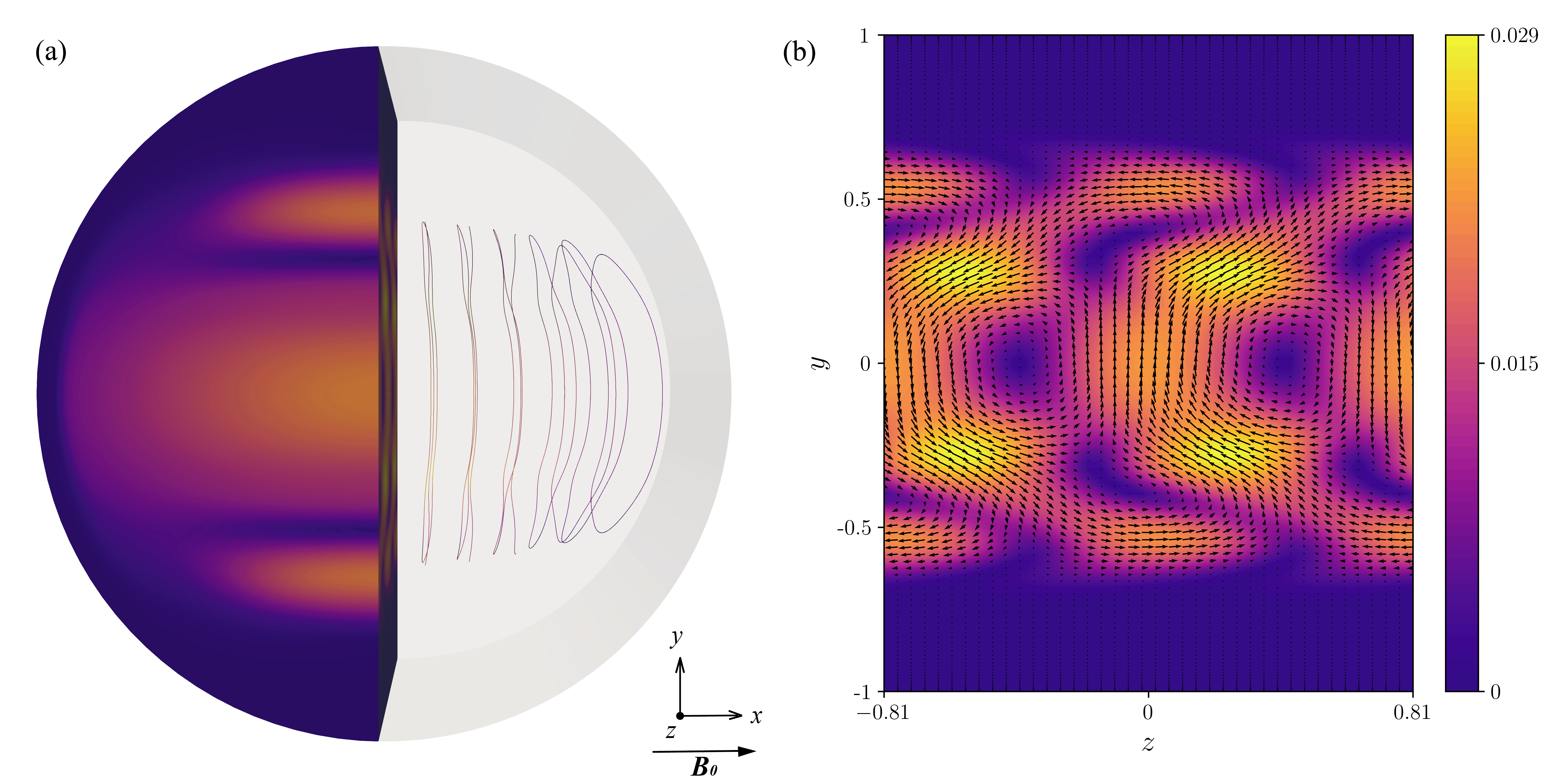}
  \caption{(a) Contours of velocity magnitude and streamlines
  of velocity field in the short-wave regime.
  (b) Corresponding $y-z$ flow field and contours of velocity magnitude. The perturbation shown
  corresponds to $\chi=0.1$, $Ha=65$, $\alpha=3.9$ and $Re=107420$.}
  \label{fig:3D-view-short-wave}
\end{figure}

In the short-wave regime, the kinetic energy is contained in the velocity
components perpendicular to the direction of the magnetic field. The
perturbations consist in counter-rotating spanwise rolls aligned with the
magnetic field (see Fig.~\ref{fig:3D-view-short-wave}). We stress again that such structures
experience much less Joule damping than the streaks characterizing the long-wave regime, and it is thus consistent that they dominate at high Hartmann numbers.

\section{Conclusions and discussion}\label{section:conclusions}
In the present study, we have analyzed the global stability of the MHD pipe
flow with a uniform transverse magnetic field using the quasi-static
approximation. For moderate Hartmann
numbers, it is found to be less stable than its non-magnetic counterpart --
Hagen-Poiseuille flow. The destabilizing effect of the magnetic field is due to
the modification of the base velocity profile. As a result, the stability
characteristics of the flow depend strongly on the Hartmann number
and the wall conductance ratio.

Our results suggest that the MHD pipe flow in a circular
insulating pipe is linearly stable up to $Re=10^6$. However, we cannot entirely
rule out the possibility of instability at higher Reynolds numbers.
In the insulating pipe, the base velocity profile only exhibits elongation in
the direction of the magnetic field, which leads to the occurrence of thin
Hartmann layers. Such Hartmann layers become unstable in the channel flow for
Reynolds numbers $Re'_c = {Re_c}/{Ha} \approx 48 311$ for sufficiently high
Hartmann numbers \cite{Lock1955, Takashima1996}. However, in the MHD pipe flow
one must account for the stabilizing effect of curved sidewalls. This has been
discussed in the context of the elliptic pipe flow in \cite{Kerswell1996}, which
also exhibits an elongated profile in one of the spanwise directions. Similarly to
the flow in a rectangular duct \cite{Tatsumi1990}, it becomes unstable to a
spanwise-modulated analogue of the most unstable perturbation in the flow
between parallel planes. However, due to the boundary curvature,
instability in the elliptic pipe occurs for much higher Reynolds numbers than in the
duct with a similar aspect ratio.

In contrast, the MHD flow in a pipe with a transverse magnetic field and electrically
conducting wall becomes unstable at sufficiently high values of the Hartmann number $Ha$
and wall conductance ratio $\chi$. In this flow regime, the base velocity profile
exhibits two sidewall maxima in the Roberts layers, which implies the possibility of
an inflection point instability \cite{Bayly1988}. Such an instability is indeed
observed and the critical Hartmann number $Ha_c$, associated with the minimum
critical Reynolds number $Re_c$ at a given value of $\chi$, is typically slightly
larger than the Hartmann number marking the emergence of sidewall velocity
overspeeds. At higher values of $Ha$, the flow stabilizes owing to stronger magnetic
damping. Two different instability regimes exist. They correspond to
$0 < \chi < 0.176$ and $\chi > 0.176$, respectively, and are mainly associated with
short or long axial wavelengths and higher or lower Hartmann numbers. In the short-wave regime, $Re_c$ attains a minimum of 107420
for $\chi=0.1$, whereas in the long-wave regime $Re_c$ monotonically with $\chi$
and converges to 45230 for $\chi\rightarrow\infty$. Moreover, the axial wavenumber
and the Hartmann number of the long-wave mode are constant when
$\chi \gtrsim 0.4$ and $\chi \gtrsim 1.9$, respectively.

The nature of the most unstable perturbations is also different in both regimes
of instability. In the long-wave regime they consist in pairs of high- and low-speed
longitudinal streaks with most of the kinetic energy concentrated in the
streamwise velocity component. These perturbations are completely antisymmetric in
cross-sections of the pipe. On the other hand, the short-wave regime is characterized
by spanwise rolls with the kinetic energy carried by the velocity components
perpendicular to the magnetic field. These perturbations are therefore
antisymmetric-symmetric in cross-sections of the pipe.
This difference in perturbation structure can be explained by the fact that the
long-wave regime corresponds to higher Hartmann numbers and that the corresponding
completely antisymmetric modes are strongly damped by Joule damping contrary to
the short-wave spanwise rolls.

Unfortunately, available experimental data about the transition in the MHD pipe
flow with a transverse magnetic field is limited.
Recent experiments conducted by Zhang \textit{et al.} \cite{Zhang2017}
aim at considering exactly the same conditions as those discussed in the present
study. However, the critical Reynolds numbers reported by the
authors for different values of $Ha$ appear to be two orders of magnitude
lower than those predicted by the linear stability theory presented here.
This discrepancy could be explained by the existence of a by-pass transition as
in Hagen-Poiseuille flow or possibly by entry effects in the flow inside the
magnetic field that could influence the transition observed in the experiments.
In this context, future works on transient growth in the MHD pipe flow with
transverse magnetic would be valuable.

\begin{acknowledgments}
The authors are grateful to A. Morozov for fruitful discussions.
This work has been carried out within the framework of 1) the EUROfusion Consortium, funded by the European Union via the Euratom Research and Training Programme (Grant Agreement No 101052200 — EUROfusion), 2) the Belgian Fusion Association and has received funding from the FPS Economy, SMEs, Self-Employed and Energy. Views and opinions expressed are however those of the author(s) only and do not necessarily reflect those of the European Union or the European Commission nor of the FPS Economy, SMEs, Self-Employed and Energy. Neither the European Union, the European Commission nor the FPS Economy, SMEs, Self-Employed and Energy can be held responsible for them.
\end{acknowledgments}

\bibliography{ref.bib}

\end{document}